\documentclass[final]{elsarticle}

\usepackage{amssymb}
\usepackage{textcomp}
\usepackage{graphicx}

\journal{Physica A}

\begin{document}

\begin{frontmatter}

\title{Towards a unified characterization of phenological phases: 
fluctuations and correlations with temperature}

\author[alabel1]{Diego Rybski}
\author[alabel1]{Anne Holsten}
\author[alabel1]{J\"urgen P. Kropp}

\address[alabel1]{
Potsdam Institute for Climate Impact Research, 
14412 Potsdam, Germany
}


\small{\today\enspace(arxiv-version 01)}

\begin{abstract}
Phenological timing -- i.e. the course of annually recurring development stages 
in nature -- is of particular interest since it can be understood as a proxy 
for the climate at a specific region; moreover changes in the so called 
phenological phases can be a direct consequence of climate change. 
We analyze records of botanical phenology and study their fluctuations
which we find to depend on the seasons.
In contrast to previous studies, where typically trends in the phenology of 
individual species are estimated, 
we consider the ensemble of all available phases and 
propose a phenological index that characterizes the influence of 
climate on the multitude of botanical species.
\end{abstract}

\begin{keyword}
phenology, phenological index, temperature, 
climate change, North Rhine-Westphalia

\PACS 

\end{keyword}

\end{frontmatter}

\section{Introduction}
\label{sec:introduction}

Phenology is a well-known concept in ecology to describe the timing of certain
periodical development stages of species throughout the year 
\cite{HudsonIL2010}. 
Developmental stages, or phases 
(e.g. flowering, fruit ripening, leaf coloring, foliation), 
have been studied over many decades in Europe using defined plant species.
This information is often used to develop phenological calendars and 
describe natural seasons \cite{BissolliMWPR2007}. 

Phenological phases are sensitive to temperature 
\cite{WaltherPCMPBFHB2002,MenzelABC2006}, and shifts of phases 
are often regarded as the first signs of a change in climate 
\cite{RosenzweigCKILMRRST2007,MenzelET2005,SchroederSH2006}.
An average earlier onset of plant phases of $3.8$\,days per $1$\,$^\circ$C 
increase over the last decades has been observed for Europe, with negative
shifts for spring and summer phases and positive shifts for fall phases 
\cite{EstrellaSM2009}.

A well-known phenological record is the cherry blossoming in Kyoto, Japan, 
which has advanced 
by $7$\,days between~1971 and~2000 \cite{AonoK2008}.
It has been shown that the flowering dates of closely related species 
in Japan have responded to climate change in a similar way 
\cite{PrimackHMR2009}. 
Nevertheless, early flowering plants deviate from this trend, showing 
larger advances due to warming than later flowering species, 
which could result in an ecological mismatch in the future.

The reaction of plants to climatic changes is non-linear and 
not uniform \cite{DoseM2006,SchleipSEM2009,SparksJKT2009}. 
It has been observed that while the correlation between air temperature and 
the onset of spring and summer plant phases is strong, 
the correlation becomes weaker for fall phases 
\cite{AbuAsabPSO2001,MenzelABC2006,WaltherPCMPBFHB2002}.
It is suggested that later in the year, other factors like
water availability, nutrition and pollution gain in importance over the 
influence of temperature \cite{EstrellaSM2009}. 
Moreover, the temporal and spatial variability of phenological
trends differs between plants and is strongest for 
spring phases \cite{MenzelSER2006}. 
Differences in the phenological response to climate warming may also result 
from locally adapted species \cite{TryjanowskiPS2006}.

Large uncertainties remain about the future development of phenological phases.
Several studies concentrate on the influence of temperature and, by assuming a
linear relation between rising air temperature and changes in the phenological
cycle, extrapolate possible future changes \cite{AhasJA2000,ChmielewskiMK2004}. 
Usually, this temperature sensitivity is analyzed by finding the 
best correlation for the preceding months of an onset date, see e.g. 
\cite{BissolliMWDRBS2005,EstrellaSM2009,FitterF2002,SparksJKT2009}.
While these previous studies concentrated on temperature responses of 
specific phases or groups of phases, no integrated approach assessing 
changes in the annual phenological cycle has been developed so far. 
We therefore propose a phenological index, which characterizes the
annual phenological cycle by taking into account 
both the shift of spring phases and the shift of fall phases 
simultaneously.
Following this approach, more general conclusions about climatic influences on
phenology can be drawn 
since more data is used, implying better statistics, and an average prospect 
is obtained.
The method is applied to the state North Rhine-Westphalia, Germany.

The paper is organized as follows. In Sec.~\ref{sec:methods} we present 
our concept of a phenological index. The data this work is based on is 
described in Sec.~\ref{sec:data}. The results of our analysis are given 
in Sec.~\ref{sec:results} in three subsections regarding fluctuations, 
the phenological index, and correlations between the index
and temperature records. In the last Section we discuss the results and 
give an outlook.

\section{Method}
\label{sec:methods}

Phenological events are referred to as phases, since they take place 
on a specific day of the year and occur at a more or less regular pace. 
For the phase~$\phi_{p,t}$, i.e. the day of the year when the 
phenological event~$p$ takes place in year~$t$, and 
the average phase over all years, $\langle\phi\rangle_p$, 
we consider the phase anomaly
\begin{equation}
\varphi_{p,t}=\phi_{p,t}-\langle\phi\rangle_p
\, ,
\label{eq:phaseanomaly}
\end{equation}
where $\langle\cdot\rangle$ denotes the average over time 
and $\langle\phi\rangle$ is defined by 
$\tan\langle\phi\rangle:=\frac{\langle\sin\phi\rangle}{\langle\cos\phi\rangle}$
\cite{TrauthM2010}, see \ref{asec:avephase}.
Accordingly, $\varphi_{p,t}$ is the anomaly record of the specific 
phenological event~$p$.
In the calculations, all phases (being originally a day of 
the calendar year) 
are transformed to the range $0\le\phi <2\pi$ by 
$\phi \rightarrow \phi\frac{2\pi}{y}$, where $y=365$ or $y=366$.

Our analysis is motivated by the following perception. 
In a year with advantageous climatological conditions, 
spring phases occur earlier than expected, e.g. as observed in an early 
flowering of Forsythia. 
In addition, fall phases occur later than expected, e.g. as seen in 
a late leaf falling of Pedunculate Oak.
In contrast, disadvantageous years lead to delayed spring phases and 
premature fall phases.
In order to capture this effect, for a given year we study the phase anomaly 
(difference between actual phase and average phase over all years) versus the 
corresponding average phase. 
In this representation, in advantageous years the anomalies of 
spring phases (located at the beginning of the year) will be negative and 
the anomalies of fall phases (at the end of the year) will be positive. 
We propose to use the statistical increase 
of the phase anomalies as a function of the average phase 
as a measure of how advantageous the climate of the 
corresponding year was for the ensemble of plants.
Thus, separately for each year, we study the parameters of a 
corresponding linear regression model for $\varphi_{p,t}$ 
against $\langle\phi\rangle_p$:
\begin{equation}
\varphi^*_{p,t}=\alpha_t\langle\phi\rangle_p+\beta_t
\, ,
\label{eq:regression}
\end{equation}
where $\alpha_t$ is the slope from the phase anomalies at year~$t$ and 
$\beta_t$ is the intercept.
From the regression, we also obtain the root mean square deviations, 
$\sigma_\varphi$, which are given by the standard deviation 
around the regression.

\begin{figure}
\begin{centering}
\includegraphics[width=0.8\textwidth]{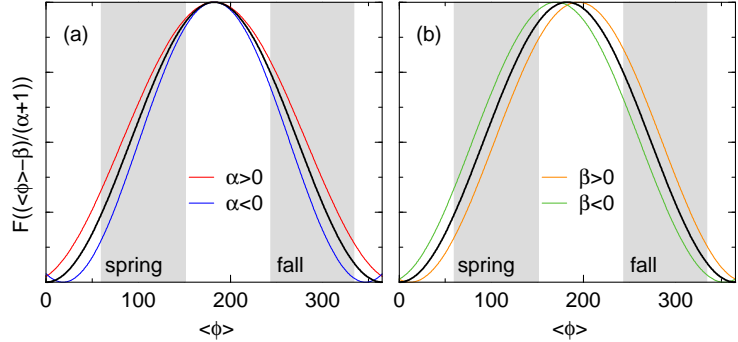}
\caption{\label{fig:illu}
Idealized cycle of advantageous and disadvantageous phenological years 
as well as premature and delayed years.
Illustration of the coefficients~$\alpha$ and~$\beta$ from 
Eq.~(\ref{eq:regression}). 
(a) Non-zero $\alpha$, which is the slope of the linear fit 
to~$\phi-\langle\phi\rangle$ versus~$\langle\phi\rangle$, 
constitutes a temporary change of the frequency. 
In the $\alpha>0$ case this leads to earlier phases in spring and 
later phases in fall.
(b) Non-zero~$\beta$, which is the intercept of the fit, constitutes 
a temporary phase shift, with overall earlier ($\beta<0$) or 
later ($\beta>0$) phases.
Compare with Eq.~(\ref{eq:phiphiavea1b}). 
For illustration we use a shifted cosine as periodic function $F$.
}
\end{centering}
\end{figure}

The linear fit, Eq.~(\ref{eq:regression}), to 
$\varphi=\phi-\langle\phi\rangle$ 
versus $\langle\phi\rangle$ provides the coefficients~$\alpha$ and~$\beta$ 
(for simplicity we skip the indices). 
Together with  Eq.~(\ref{eq:phaseanomaly}) one obtains 
(eliminating $\varphi$)
\begin{equation}
\phi=\langle\phi\rangle(\alpha+1)+\beta
\, ,
\label{eq:phiphiavea1b}
\end{equation}
which shows that $\alpha$ corresponds to a temporary change of frequency. 
Figure~\ref{fig:illu}(a) illustrates that a positive slope $\alpha$ is 
related to a low frequency anomaly causing early phenological phases in spring 
and late phenological phases in fall (see also Fig.~\ref{fig:bsp196090}).
In the same way, $\beta$ corresponds to a temporary phase shift, 
as illustrated in Fig.~\ref{fig:illu}(b) -- all phases appear before or 
after the average.

We understand phenological processes to be triggered by an 
annual cycle that is a compound of all relevant climatological features. 
In general, such a cycle is unknown but we think of it as illustrated in 
Fig.~\ref{fig:illu}. 
Once it passes a certain threshold, and its derivative has the right sign, 
such as positive for spring or negative for fall, 
a specific plant is activated and a phenological phase takes place, 
e.g. flowering in spring.

In~\ref{asec:phenocyclealphe} we show that the slope~$\alpha$ is also 
associated with an increased or decreased cycle in such a way that the 
integral over an idealized annual cycle~$C$ is approximately proportional 
to~$\alpha$, as suggested by Fig.~\ref{fig:illu}(a).

\section{Data}
\label{sec:data}
The study region, North Rhine-Westphalia (NRW), is the most populous state of
Germany ($\approx$ $18$\,million residents in 2008; 
$34,070$~km$^2$ total area). 
Two types of landscapes can be found in NRW: the North German lowlands 
with an elevation just a few meters above sea level, and the North German low 
mountain range with elevations of up to $850$\,m.
The lowlands comprise the Rhine-Ruhr Area which is one of the largest 
metropolitan areas worldwide. 
These landscape features are also expressed by distinct types of climate. 
While in the lowlands the mean annual temperature is 
$10$\,$^\circ$C with an annual mean precipitation of $620$\,mm, 
in the mountainous regions the mean temperature is $5$\,$^\circ$C and 
an annual mean precipitation of up to $1,500$\,mm 
is common as measured between the years 1961-1990
\cite{OesterleWG2006}.

Onset dates of numerous phenological phases have been collected in Germany by
the German Weather Services (DWD) for the past decades. 
Observations are carried out two to three times in a week, 
which determines the temporal accuracy of the dataset. 
Since 1951 data for over $159$~phases has been observed at around 
$660$~stations in NRW. 
Due to incomplete datasets, especially before 1970, 
we have reduced the number of stations to those providing 
sufficient data for our purposes over the whole period 
from~1951 to~2006 (see \ref{asec:phenolist}). 
As agricultural phases are strongly influenced by agricultural practices 
as well as breeding, and show a weaker relation to 
temperature changes \cite{MenzelVESD2006}, 
they were not considered in this study.  
Thus, we analyzed time series data of 17 meteorological and phenological 
stations in NRW for $75$~phases for the period of 1951-2006 
(cf. Fig.~\ref{fig:map}).
In order to investigate the effect of temperature on phenology, 
annual mean temperature records from the nearest climate station to each 
phenological station were further taken into account. 
Temperature records are based on observational data of the DWD and 
were partly interpolated \cite{OesterleWG2006}.
In \ref{asec:phenolist} we list the phenological phases and 
climatological stations.

\section{Results}
\label{sec:results}

\subsection{Fluctuations}
\label{ssec:fluct}

\begin{figure}
\begin{centering}
\includegraphics[width=0.8\textwidth]{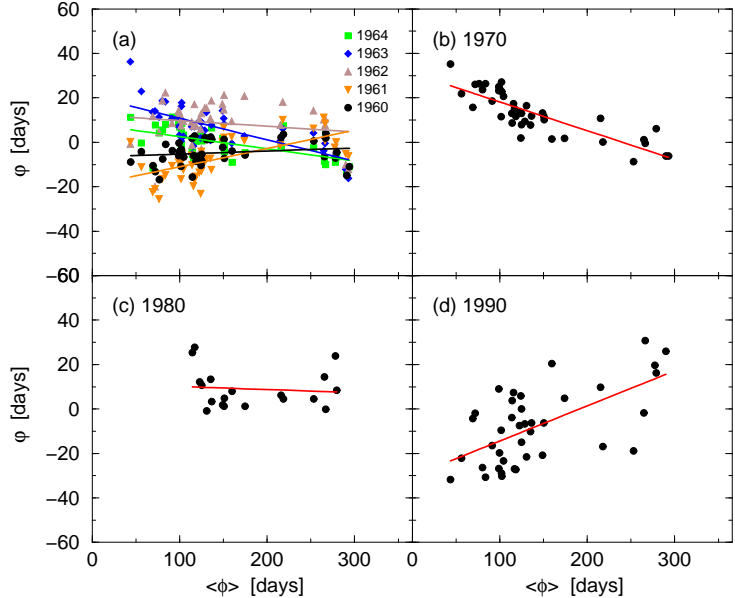}
\caption{\label{fig:bsp196090}
Examples of phase anomalies versus average phase in the case of 
D\"ulmen near M\"unster in North Rhine-Westphalia for the years 
(a)~1960-1964, (b)~1970, (c)~1980, and (d)~1990.
The filled symbols represent the various phenological phases and 
the solid line is a linear fit through the data by least squares.
}
\end{centering}
\end{figure}

In Figure~\ref{fig:bsp196090} we show examples of 
$\varphi=\phi-\langle\phi\rangle$ versus $\langle\phi\rangle$, 
namely for the years 1960-1964, 1970, 1980, and 1990 at the station 
D\"ulmen.
During winter, i.e. approx. $\langle\phi\rangle<50$ and 
$\langle\phi\rangle>300$, no phenological activity is recorded. 
While in 1960 [Fig.~\ref{fig:bsp196090}(a)] the phenological phases appear 
more or less as in average, in 1961 spring phases occurred prematurely.
In 1962 all phases were delayed and in 1963 and 1964 the spring phases only.

In 1970 [Fig.~\ref{fig:bsp196090}(b)] spring 
phases occured late ($\varphi>0$) leading to a negative slope~$\alpha$.
In 1980 [Fig.~\ref{fig:bsp196090}(c)] less phases were recorded but 
on the basis of the available data it seems to have been a rather normal 
year. 
In 1990 [Fig.~\ref{fig:bsp196090}(d)] the early phases appear prematurely 
($\varphi<0$) indicating good conditions in spring.

Next we want to address how strongly the phenological phases fluctuate.
In order to quantify these fluctuations we use the Rayleigh measure
\cite{RosenblumPKST01,TrauthM2010}:
\begin{equation}
\sigma_\phi=\sqrt{\langle\cos\phi\rangle^2+\langle\sin\phi\rangle^2}
\, .
\label{eq:phephafluc}
\end{equation}
Here $\langle\cdot\rangle$ is the average over time, 
separately for each phenological plant. 
If~$\phi$ is spread uniformly over the period 
then~$\sigma_\phi$ is close to~$0$ 
because the averages of the trigonometric functions are very small.
If~$\phi$ is fixed then~$\sigma_\phi$ is~$1$. 
Thus, values close to~$0$ indicate large fluctuations and values 
close to~$1$ small ones.
We use this quantity since the standard deviation of an angle is not 
well defined.
We would like to remark that $\sigma_\phi$ is independent of the regressions 
in Fig.~\ref{fig:bsp196090}.

\begin{figure}
\begin{centering}
\includegraphics[width=0.8\textwidth]{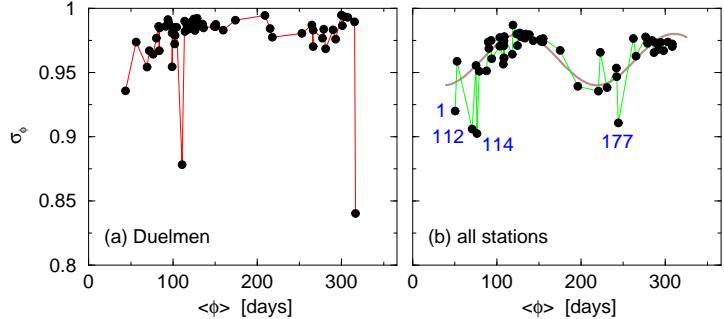}
\caption{\label{fig:bspavestd}
Fluctuations according to Eq.~(\ref{eq:phephafluc}) of phenological phases 
(a) in the case of D\"ulmen in North Rhine-Westphalia and 
(b) for all stations. 
In both cases the average is over all available years.
Small~$\sigma_\phi$ correspond to large fluctuations. 
The solid line in the background of (b) illustrates the wavy pattern of 
the fluctuations 
(functional form $\propto\sin\langle\phi\rangle\cos\langle\phi\rangle$).  
The numbers indicate some phenological phases that are spread 
relatively strongly.
1: Hazel, {\em Corylus Avellana}: flowering; 
112: European Alder, {\em Alnus Glutinosa}: flowering; 
114: Cornel Cherry, {\em Cornus Mas}: flowering; and
177: Wild Brier, {\em Rosa Canina}: fruit ripening.
}
\end{centering}
\end{figure}

Figure~\ref{fig:bspavestd} shows the spreading~$\sigma_\phi$ versus 
the average phase~$\langle\phi\rangle$. 
The result for the example from Fig.~\ref{fig:bsp196090} is depicted in 
Fig.~\ref{fig:bspavestd}(a).
Two phenological phases have small values of~$\sigma_\phi$ and 
accordingly large spreading -- 
which is due to measurement errors. 
Apart from that, most phases show $\sigma_\phi>0.95$ and only 
early phases exhibit larger fluctuations (smaller~$\sigma_\phi$), 
compare with Fig.~\ref{fig:bsp196090}(a).

In contrast, the~$\sigma_\phi$ obtained from all stations 
[Fig.~\ref{fig:bspavestd}(b)] look smoother and four phenological 
phases have rather small $\sigma_\phi$-values (large spreading). 
In general, a kind of wave pattern can be observed 
and is illustratively traced in Fig.~\ref{fig:bspavestd}(b): 
Spring phases exhibit larger fluctuations, 
early summer phases smaller ones, 
late summer phases again larger fluctuations, 
and fall phases again small ones. 
Calculating standard deviations, similar patterns have been found 
for $35$~plant phases and $29$~butterfly phases \cite{MenzelSER2006}.
Assuming a wave (see Fig.~\ref{fig:illu} and~\ref{asec:phenocyclealphe}) 
those phases with small fluctuations coincide with large slopes 
(or small negative slopes) of an idealized phenological cycle. 

Deviations from the curve could result from measurement 
inaccuracies, since for some phases (e.g. fruit ripening) the exact 
onset date is difficult to determine. 
Another reason could be that those phenological phases with 
small fluctuation are triggered by a sharp change while the others with 
larger fluctuations typically occur in seasons when the 
trigger is not as sharp. 
In other words, small deviations of the phenological cycle barely 
influence those phases that in average occur when the idealized cycle has 
a large slope.
Contrariwise, small deviations of the phenological cycle do affect phases 
that in average occur when the idealized cycle has a small slope.

\subsection{Phenological index}
\label{ssec:phenoindex}

\begin{figure}
\begin{centering}
\includegraphics[width=0.8\textwidth]{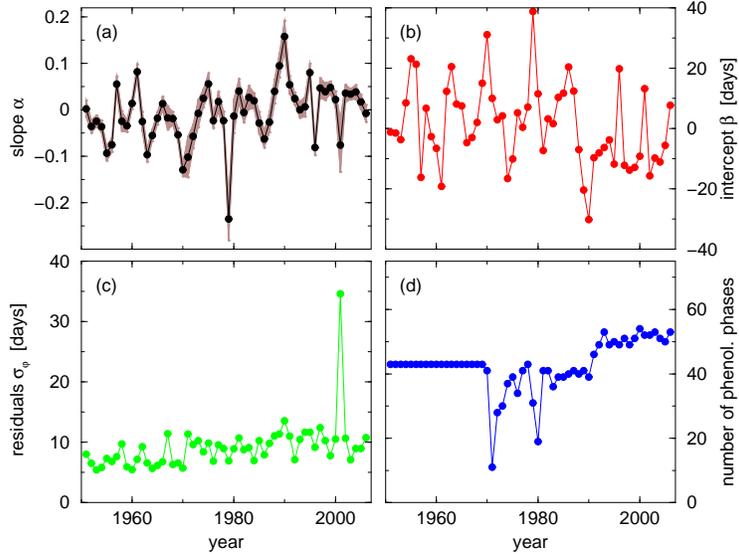}
\caption{\label{fig:bspslpinter}
Values obtained for the phase anomaly fit Eq.~(\ref{eq:regression}) as 
illustrated in Fig.~\ref{fig:bsp196090} for the years 1951-2006 
(D\"ulmen).
(a)~slope $\alpha$ (pheno-index), 
(b)~intercept $\beta$, 
(c)~root mean square deviations from the fit $\sigma_\varphi$, and 
(d)~number of phenological phases used for each year.
The brownish area in the background of (a) corresponds to the 
standard error.
}
\end{centering}
\end{figure}

Systematically applying linear regressions to 
$\varphi=\phi-\langle\phi\rangle$ versus $\langle\phi\rangle$ 
of the example station D\"ulmen 
(solid lines in Fig.~\ref{fig:bsp196090})
we obtain a set of quantities in Fig.~\ref{fig:bspslpinter}, 
plotted against the corresponding year.
In Fig.~\ref{fig:bspslpinter}(a) and~(b) we show the two fit coefficients, 
namely the slope, $\alpha$, and the intercept, $\beta$, respectively. 
As pointed out in Sec.~\ref{sec:methods}, the former indicates 
how advantageous a year is. 
We consider it as a phenological index (pheno-index). 
As can be seen, $\alpha$ fluctuates from year to year roughly in the range 
$-0.2<\alpha<0.2$.

Two additional quantities of interest are depicted in 
Fig.~\ref{fig:bspslpinter}(c) and~(d). 
The root mean square deviations from the fit in Fig.~\ref{fig:bsp196090}, 
$\sigma_\varphi$, capture how uniform the annual cycle is or 
how homogeneously the phenological phases respond to the climate variations. 
We find that except from an outlier in the year 2001 
(due to a measurement error) 
the residuals are stable with values around or below $10$\,days.
Remarkably, the outlier does not seem to affect much the values of 
$\alpha$ and $\beta$ in 2001.
Figure~\ref{fig:bspslpinter}(d) shows the number of phenological phases 
considered in the specific years; this is the same number of points appearing 
in the panels of Fig.~\ref{fig:bsp196090}.
Somehow -- for the example station -- up to 1970 constantly 43 phases 
were recorded per year. 
In 1971 only 11 values are considered but still $\alpha$ and $\beta$ 
seem to have reasonable values, supporting the robustness of the 
approach.

\begin{figure}
\begin{centering}
\includegraphics[width=0.8\textwidth]{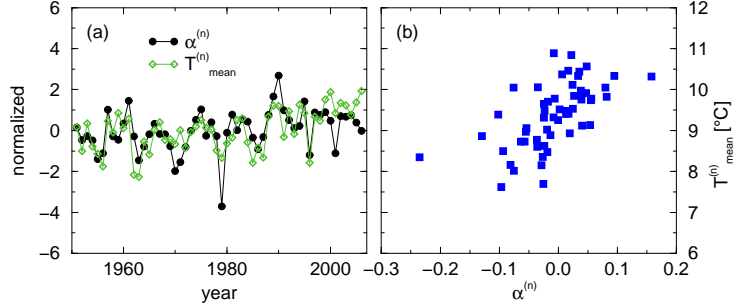}
\caption{\label{fig:bspcorr}
Correlations between pheno-index and annual mean temperature for 
the years 1951-2006 (D\"ulmen).
(a)~Comparison of annual mean temperature $T_{\rm mean}$ (open diamonds) 
and pheno-index~$\alpha$ (filled circles).
(b)~Scatter plot of~$T_{\rm mean}$ versus~$\alpha$ for all years. 
For a better comparison, the records are normalized according to 
Eq.~(\ref{eq:normtemp}) and Eq.~(\ref{eq:normalph}), respectively.
}
\end{centering}
\end{figure}


\subsection{Correlations between phenological phases and 
temperature records}
\label{ssec:corrphaenotemp}

It is known that the temperature is an important climatological element 
influencing the phenological timing, in particular at springtime 
\cite{WaltherPCMPBFHB2002}.
Next we want to inspect, how the pheno-index (slope~$\alpha$) 
is related to the mean annual temperature $T_{\rm mean}$. 
Figure~\ref{fig:bspcorr}(a) shows both, $\alpha$ as well as $T_{\rm mean}$ 
measured at the closest climatological station, nearby Billerbeck, 
which is situated less than $20$\,km from D\"ulmen. 
In order to compare the two quantities, we have normalized both records to 
zero average and unit standard deviation:
\begin{equation}
T_{\rm mean}^{\rm (n)}=(T_{\rm mean}-\langle T_{\rm mean}\rangle)/\sigma_T
\label{eq:normtemp}
\end{equation}
as well as 
\begin{equation}
\alpha^{\rm (n)}=(\alpha-\langle\alpha\rangle)/\sigma_\alpha
\label{eq:normalph}
\, .
\end{equation}
We find a fair agreement between the course of both quantities. 
However, from 1999 onwards the normalized temperature values are above the 
normalized pheno-index.
By definition, due to continuity reasons, $\alpha$ cannot systematically 
deviate from zero.
Thus, further research is required to reveal if this is a systematic 
deviation or within the statistical fluctuations.

In Figure~\ref{fig:bspcorr}(b) the quantities~$\alpha$ and~$T_{\rm mean}$ 
are plotted against each other for each year. 
The correlation coefficient for this example is~$0.60$, 
which is a satisfying result considering the noisy data of 
Fig.~\ref{fig:bsp196090}.

\begin{figure}
\begin{centering}
\includegraphics[width=\textwidth]{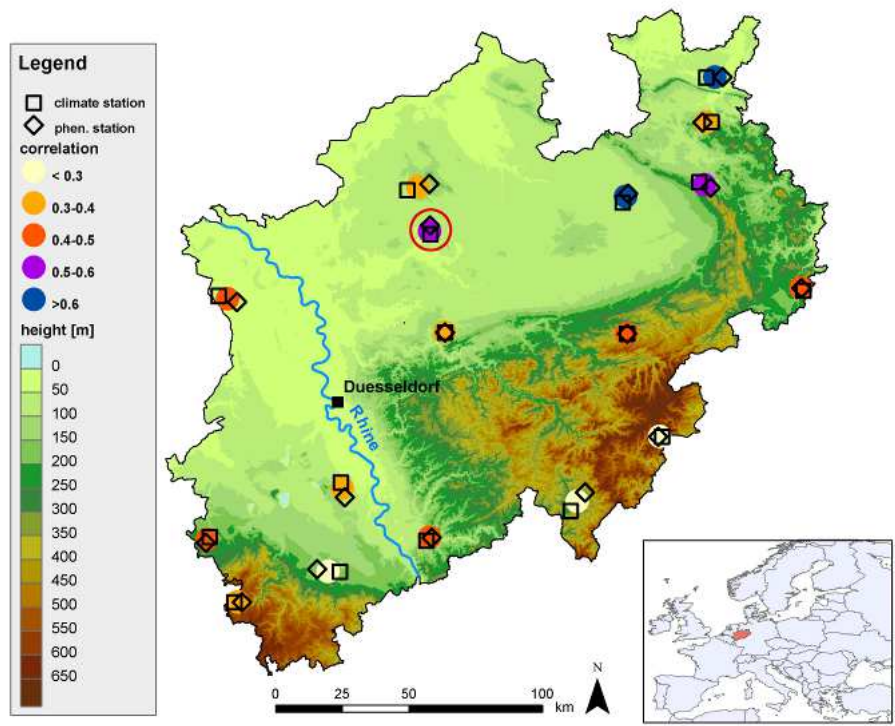}
\caption{\label{fig:map}
Area under investigation, North Rhine-Westphalia, and correlations between 
annual mean temperature and pheno-index.
Since, in general, phenology (open diamonds) and temperature (open squares) 
are not measured at the same site, we identify the closest pairs and 
attribute the correlation value as a color-coded circle to the 
center between the two.
The exemplary station D\"ulmen illustrated in the previous Figures is 
indicated by the red solid line circle.
}
\end{centering}
\end{figure}

Figure~\ref{fig:map} exhibits the area under investigation, 
North Rhine-Westphalia. We identify for each phenological station the 
closest temperature station and calculate the correlation value between 
the pheno-index and the associated annual mean temperature record. 
The locations of the stations and the resulting correlation values are 
depicted in the map. The correlations vary between~$0.28$ 
and~$0.64$ and do not seem to show any systematic dependence on 
the position, indicating that micro-scale spatial 
climatological conditions may be dominating the pheno-index.
In addition, the correlation value of $\alpha$ and $T_{\rm mean}$ does not 
significantly depend on the amount of phenological data of the stations.

For several reasons we do not expect much larger correlation coefficients 
for individual phenological stations. 
Certainly, the phenological phases are also influenced by other factors, 
in particular precipitation or sunshine duration. 
Precipitation has weak spatial correlations and could be responsible 
for micro-scale influences. 
Hence, including further information of this kind could improve the 
correlations.
In addition, the spring phases are influenced by the beginning of the year 
and possibly by the end of the year before but certainly not by the end of 
the actual year. Thus, considering seasonal values or 
taking into account at least part of the 
previous year could also lead to stronger correlations. 
In addition, the fall phases are known to react in a less pronounced way 
to temperature 
\cite{AbuAsabPSO2001,MenzelABC2006,WaltherPCMPBFHB2002} and could contribute 
to noise leading to reduced correlation values.

\section{Discussion and Conclusions}
\label{sec:discussion}

In summary, we analyze phenological records, characterize the 
fluctuations of the phases, and introduce a phenological index.
We find that the spring and late summer phases exhibit the 
largest fluctuations while the early summer and fall phases exhibit the 
smallest fluctuations. This may be related to the derivative of 
the annual cycle, such as of temperature.
By plotting (for each individual station) the phase anomaly 
against the average of 
each phase and applying a linear regression through all phases, 
we obtain a measure for how advantageous a specific year was for the 
ensemble of flora at the corresponding site. 
The slope represents a temporary change 
of frequency and the intercept a temporary phase shift.
In addition, we show that the slope of such a fit is approximately 
proportional to the integral over the idealized annual 
phenological cycle.
The advantage of our approach is that it characterizes the multitude of 
climatological factors influencing the entire phenological ensemble 
considered. 
It smoothes the volatile phenological data to 
a combined index which helps to detect and quantify impacts on life and 
life cycles.
It can be applied even when records are incomplete, 
something which can cause problems when phenological phases 
are considered individually.
Generally, a similar index can also be calculated for the 
phenology of fauna, but this is a task we leave for future studies. 

We compare the pheno-index with the annual mean temperature and 
find some agreement. The correlation value for various stations in 
the area of investigation varies between~$0.28$ and~$0.64$ whereas 
from this study we do not find any systematic dependence of the 
correlations on the location.
We conclude that additional factors influence the pheno-index.
The response of phenological phases to changes in temperature 
has been found to vary spatially, for example being 
stronger in more northerly latitudes \cite{EstrellaSM2009}. 
Further, a more pronounced spring advance has been described for 
maritime western and central Europe compared to the 
continental east \cite{AhasJA2000}.
Temperature has been identified as the main driver of spring phases, 
followed by the photoperiod length. Other factors such as precipitation, 
nutrient availability and soil properties showed only minor effect in 
comparison to temperature \cite{BadeckBBDLSS2004}. 
Further research is needed in order to figure out how our approach 
relates to these previous findings.

Climate change induced shifts in phenology could disrupt the chain between
pollinator and plants \cite{MemmottCWP2007}.
Phenological plant phases are key stages of plant development -- 
changes in their timing might influence other species. 
Thus, alterations of phenological phases could disrupt 
interactions among species, e.g. within food webs \cite{MenzelSER2006}, 
see also \cite{CamachoGA2002}. 
Evidence from various species indicate an insufficient rate of 
phenological adaptation concerning food webs to 
changing climatic conditions \cite{VisserB2005}.

It has been found that correlations between air temperature and 
fall phases are less pronounced 
\cite{AbuAsabPSO2001,MenzelABC2006,WaltherPCMPBFHB2002}.
Thus, the application of the ideas of this work could lead to a better 
consideration of fall phases within phenological analysis.
There is considerable interest on how phenology will be affected
by climate change, particularly in the context of ecology or 
agriculture. 
The phenological index could be used as the basis of projections obtained from 
climate models to project the changes in phenology.

\appendix

\section{Calculating the average phase $\langle\phi\rangle$}
\label{asec:avephase}

We want to calculate the average phase, $\langle\phi\rangle$, 
from a set of angles $\phi_i$. 
In order to account for the cyclicality of the phase, we do not simply 
average the phases, but consider the Euler relation, 
${\rm e}^{{\rm i}\phi}=\cos\phi+{\rm i}\sin\phi$, 
average the sine and the cosine separately, and use 
the relation $\tan\theta=\frac{\sin\theta}{\cos\theta}$. 

However, writing the inverse, 
$\langle\phi\rangle = 
\arctan\left(\frac{\langle\sin\phi\rangle}{\langle\cos\phi\rangle}\right)$, 
is not precise since the $\arctan$-function does not take into account 
the signs of numerator and denominator. 
Therefore, most programming languages provide the 
two argument function ${\rm atan2}$, which properly calculates the angle, 
\begin{equation}
\langle\phi\rangle = 
{\rm atan2}\left(\langle\sin\phi\rangle,\langle\cos\phi\rangle\right)
\, .
\end{equation}

\section{Relation between the pheno-index~$\alpha$ and the anomaly of the phenological cycle}
\label{asec:phenocyclealphe}

Although, the annual cycle of phenological advantage 
basically can have any periodic shape, 
we assume such a cycle has the form of a sine-wave 
(for simplicity we skip the indices~$p$ and $t$):
\begin{equation}
C(\phi)=A\sin(\phi+\lambda)+B
\, ,
\label{eq:idcycle}
\end{equation}
where $A$ is the amplitude, $\phi=\nu t$ the phase (frequency $\nu$), 
$\lambda$ the phase shift, and $B$ an offset. 
We would like to remark that it would be more meaningful to express $\phi$ 
as a function of $C$, since the annual cycle triggers the phenological phases.
Due to climate fluctuation, the cycle~$C$ deviates from the average annual 
cycle within a year as well as from year to year.
Since such an idealized cycle is unknown, we study the phenological signals.

Here we show that the slope $\alpha$ is associated with an increased or 
decreased cycle in such a way that the integral over $C$ is approximately 
proportional to $\alpha$, as suggested by Fig.~\ref{fig:illu}(a).

The integral over one period of the \emph{average} annual cycle 
vanishes when we drop the offset
\begin{equation}
\int_{-\pi}^\pi\!
\left[A\sin(\langle\phi\rangle+\langle\lambda\rangle)\right]
{\rm d}\langle\phi\rangle=0
\, .
\label{eq:acintzero}
\end{equation}
The quantities $\phi$ and $\lambda$ are spread around 
$\langle\phi\rangle$ and $\langle\lambda\rangle$, respectively 
(assuming $A={\rm const.}$), 
and are in general different 
from the averages.
Using Eq.~(\ref{eq:phiphiavea1b}) in (\ref{eq:idcycle}) we express the 
integral over $C$ as
\begin{eqnarray}
& &
\int_{-\pi}^\pi\!
\left[A\sin(\langle\phi\rangle(\alpha+1)+\beta+\lambda)\rangle\right]
{\rm d}\langle\phi\rangle 
\nonumber \\
& = & 
\left[-\frac{A}{(\alpha+1)}\cos(\langle\phi\rangle(\alpha+1)+\beta+\lambda)\right]_{-\pi}^\pi 
\nonumber \\
& = & -\frac{A}{(\alpha+1)}
\left[\cos(\pi(\alpha+1)+\beta+\lambda)-\cos(-\pi(\alpha+1)+\beta+\lambda)\right] 
\nonumber \\
& = & -\frac{A}{(\alpha+1)}
[\cos(\pi(\alpha+1))\cos(\beta+\lambda)-\sin(\pi(\alpha+1))\sin(\beta+\lambda)
\nonumber \\
& & { } -\left( \cos(-\pi(\alpha+1))\cos(\beta+\lambda)-\sin(-\pi(\alpha+1))\sin(\beta+\lambda) \right) ] 
\nonumber \\
& = & \frac{2A}{(\alpha+1)} \sin(\pi(\alpha+1))\sin(\beta+\lambda)
\end{eqnarray}
Since $\beta+\lambda\approx -\pi/2$ 
(in order to match the seasons within the calendar year) 
the second term is $\sin(\beta+\lambda)\approx -1$. 
For $\alpha$ close to $0$, the first term goes like $-\pi\alpha$ and 
assuming $\alpha+1\approx 1$ one obtains
\begin{eqnarray}
\int_{-\pi}^\pi\!
\left[A\sin(\langle\phi\rangle(\alpha+1)+\beta+\lambda)\rangle\right]
{\rm d}\langle\phi\rangle
{} & {} \approx {} & {} 
2\pi A\alpha \\
{} & {} \sim {} & {}
\alpha 
\, . 
\end{eqnarray}
Therefore, we conclude that $\alpha$, as the regression slope to 
$\phi-\langle\phi\rangle$ versus $\langle\phi\rangle$, 
is a measure for the anomaly of the phenological cycle with 
respect to early spring phases and late fall phases and vice versa.
However, the unchanged maximum in Fig.~\ref{fig:illu}(a) is not very 
realistic. Having in mind a temperature change, one would expect an 
overall vertical shift of the cycle and accordingly rather an anomaly 
reflected in the offset $B$.
Nevertheless, since there are few phenological phases in winter 
(where no events are measured) and 
since the analysis is performed statistically, $\alpha$ can still be 
considered as a measure for the in- or decrease of the phenological cycle.

\section{Details on the phenological data}
\label{asec:phenolist}

The observational program of wild plants includes the following codes:
1-20, 64-74, 112-135, 175-178, 213-228, which are listed at 
http://www.dwd.de $>$ Climate + Environment $>$ Phenology $>$ 
Observation programme $>$ Wild plants.
In order to have sufficient statistics, we filter the phenology data 
according to the following criteria: 
(i) phenological phases with at least $3$~entries for one station, 
(ii) years with at least $3$~pairs of 
$\langle\phi\rangle$,~$\varphi$, and 
(iii) stations with at least $30$~years of data, whereas the presence of 
the years~1951 and~2006 is required.
The phenological and associated temperature stations are listed in 
Tab.~\ref{tab:stations}. 
Daily temperature records have been averaged to annual resolution.
Missing temperature data has been interpolated \cite{OesterleWG2006}.

\begin{table}
\caption{Phenological and associated temperature stations}
\label{tab:stations}
\begin{center}
\footnotesize
\begin{tabular}{| l l | l l |}
\hline
\multicolumn{2}{|c|}{phenological stations} &
\multicolumn{2}{|c|}{climate stations} \\
\hline
ID & location & PIK-ID & location \\
\hline\hline
52334110 & Kevelaer, Kleve & 19183 & Weeze-Hees \\
53331130 & Z\"ulpich, Euskirchen & 19006 & Euskirchen \\
53341120 & Frechen, Erftkreis & 19107 & Pulheim-Brauweiler \\
53371170 & Hennef, Rhein-Sieg-Kreis & 19113 & Hennef \\
54110000 & Aachen (DWD), kreisfreie Stadt Aachen & 19004 & Aachen \\
54352120 & Imgenbroich, Aachen & 19119 & Monschau \\
55342130 & Billerbeck, Coesfeld & 19175 & Coesfeld \\
55344110 & D\"ulmen, Coesfeld & 19177 & Billerbeck \\
57331490 & Heidenoldendorf, Lippe & 20339 & Lage, Kr.Lippe-Hoerste \\
57359110 & Exter, Herford & 15208 & Vlotho-Valdorf \\
57391110 & Minden, Minden-L\"ubbecke & 15182 & Minden-Hahlen \\
57412130 & B\"uhne, H\"oxter & 20031 & Borgenb \\
57421110 & G\"utersloh, G\"utersloh & 20022 & Guetersl \\
58326170 & Warstein, Hochsauerlandkreis & 20259 & Warstein \\
58397360 & Obernetphen, Soest & 20013 & Siegen \\
58422410 & Wunderthausen, Siegen-Wittgenstein & 20291 & Berleburg, Bad-Wunderthausen \\
59230000 & Witten-Stockum, Ennepe-Ruhr-Kreis & 19221 & Witten-Stockum \\
\hline
\end{tabular}
\end{center}
\end{table}

\section*{Acknowledgments}
This work was supported by the Ministry of the Environment, Regional Planning 
and Agriculture of North Rhine-Westphalia and by the ESPON Climate project 
(partly funded by the European Regional Development Fund).
We also appreciate financial support by 
BaltCICA (Baltic Sea Region Programme 2007-2013).
We are thankful to Kirsten Zimmermann from DWD (German Meteorological Service) 
for assistance with the data, to the various volunteers for recording 
the phenological phases, and to Dominik Reusser, Alison Schlums, as well as 
Carsten Walther for fruitful discussions.

\section*{References}
\bibliography{phaeno}

\end{document}